%% file: Two-sided.tex
\documentclass[11pt]{article}

\usepackage{fullpage}

\usepackage{natbib}
 \bibpunct[, ]{(}{)}{,}{a}{}{,}%

\usepackage{amsmath,amssymb,amsfonts,mathrsfs, mathtools, bm, bbm, dsfont, mathrsfs}
\usepackage{graphicx}
\usepackage{ifthen}
\usepackage{multirow, multicol}
\usepackage{float,framed}
\usepackage{subeqnarray, array}
\usepackage{enumitem}
\usepackage{xcolor}
\usepackage{caption}
\usepackage{booktabs}
\usepackage{float}
\usepackage[caption=false]{subfig}
\usepackage{natbib}
\usepackage[font=small]{caption}
\usepackage{graphicx}
\newcommand{\RSI}{{\sf RSI}}

\newcommand{\LI}{{\sf LI}}

\usepackage[breaklinks, colorlinks, linkcolor=purple, anchorcolor=purple, citecolor=purple, urlcolor=purple]{hyperref}

\input{header}

\usepackage{palatino}

\begin{document}
%%%%%%%%%%%%%%%%

%\begin{document}

\title{\bf A Scalar-Parameterized Mechanism For Two-Sided Markets}
\author{Mariola Ndrio \qquad Khaled Alshehri \qquad Subhonmesh Bose} %\\ \\ Working paper}

%\KEYWORDS{convex relaxation, locational marginal prices, revenue adequacy, efficient market equilibrium}

\maketitle

\begin{abstract}
We consider a market in which both suppliers and consumers compete for a product via scalar-parameterized supply offers and demand bids. Scalar-parameterized offers/bids are appealing due to their modeling simplicity and desirable mathematical properties with the most prominent being bounded efficiency loss and price markup under strategic interactions. Our model incorporates production capacity constraints and minimum inelastic demand requirements. Under perfect competition, the market mechanism yields allocations that maximize social welfare. When market participants are price-anticipating, we show that there exists a unique Nash equilibrium, and provide an efficient way to compute the resulting market allocation. Moreover, we explicitly characterize the bounds on the welfare loss and prices observed at the Nash equilibrium.
\end{abstract}

\section{Introduction}
The distinction between consumers and producers in marketplaces is increasingly fading. In the retail electricity sector, increased consumer participation---either as generation suppliers or price-responsive demanders---is driving the emergence of a digital platform marketplace where end-use customers can engage in transcations coordinated via a central entity or market manager as proposed in \citet{Car}. Similar digital marketplaces have emerged in the areas of ride-sharing (Uber, Lyft), lodging (Airbnb), online retail and trading auctions (Amazon, Ebay) etc. A common feature of these multi-sided marketplaces is a collection of agents who can take up the mantle of being suppliers or consumers while the market clears through a centralized mechanism, often operated by a market manager. Motivated by these transformations,
in this paper we study a two-sided market with a finite number of suppliers that compete to supply a product to price-responsive consumers. Our focus is on uniform price markets that clear through a centralized mechanism that sets a single per-unit price on the product for all participants. Every consumer (supplier) expresses her willingness to buy (offer) via a demand bid (supply offer) that fully characterizes her demand (supply) quantity at a given market price. 
We investigate the following market design question: \emph{What is the right mechanism that allows market actors sufficient flexibility to declare their willingness to offer/buy such that it yields efficient allocations, i.e., an allocation that maximizes social welfare?} 

The seminal work by \citet{Klemperer} demonstrated that in the absence of uncertainty there exist an enormous multiplicity of equilibria in supply functions. Hence, there is a need to resort to stylized offer/bid functions that appropriately restrict the family of supply offers and demand bids which the market actors are allowed to utilize. The well-known Bertrand and Cournot competition models are examples of simple (degenerate) supply offer strategies in markets with uniform prices. However, the Bertrand model typically assumes that each participant is willing to supply the entire demand, which may not be satisfied in a number of cases. Variations of the Bertrand model with capacity constraints have been proposed, however, in such settings pure Nash equilibria may not exist as shown in \citet{Sh:59}. The Cournot model has a number of appealing properties when studying oligopolies in markets with relatively high demand elasticity. However, when demand elasticity is low Cournot competition may exhibit arbitrarily high welfare loss (see \citet{Day}). Furthermore, pure quantity or price competition cannot adequately represent markets with more complicated offer structures. An example of such markets are day-ahead wholesale electricity markets that operate either as pools or power exchanges. In these markets, power producers submit offers to supply varying quantities at succesively higher prices and the demand side specifies the quantity willing to purchase at succesively lower prices. 
Linear supply functions is another candidate family of functions to model strategic interactions among suppliers. However, the work of \citet{Baldick} illustrates that it is not straightforward to incorporate capacity constraints into linear supply offers. Moreover, arbitrary high efficiency loss at the Nash equilibrium is possible, particularly when suppliers have highly heterogeneous cost functions (see \citet{nali}).

In this work, we restrict our attention on a specific family of supply offers and demand bids, referred to as scalar-parameterized supply functions, studied by \citet{Johari:03} and \citet{Johari:11} in markets with inelastic supply and demand respectively. The specific family of offer/bid functions allows market actors to have one-dimensional action spaces, when faced with a single market price. Such market mechanisms are simple to implement and are considered to be fair among market participants. Moreover, the work of \citet{Johari:04} and \citet{Johari:11} showed that such supply offers possess a number of attractive properties including bounded Price of Anarchy and price markups at the Nash equilibrium. The family of supply functions considered here is a capacitated version of those introduced by \citet{Johari:11} that have been studied by \citet{Xu} and \citet{Lin} under perfectly inelastic demand. Such supply functions prohibit situations where firms can offer in the market beyond their means.

In this paper, we aim to study the most general setting: a two-sided market where both supply and demand compete through supply offers and demand bids, which we present in Section \ref{sec:model}.
In sections \ref{sec:pc} and \ref{sec:Nash}, we characterize the market outcome in situations when all market actors are (i) pure price-takers and (ii) price-anticipating. We show that under perfect competition our market mechanism yields allocations that maximize social welfare. When both sides of the market are price-anticipating, the misrepresentation of private information has the potential to induce market allocations that are suboptimal to the efficient outcome. However, our analysis in Section \ref{sec:efficiency} indicates that both the welfare loss and the price markup at the Nash equilibrium are bounded. Numerical experiments in Section \ref{sec:numerics} illustrate the main insights of the analysis. Section \ref{sec:conclusion} concludes the paper. All proofs are provided in the Appendix.

\emph{Notation:} Let $\Rset$ denote the set of real numbers and $\mathbb{R}_{+}$ the set of non-negative real numbers. Denote the transpose of a vector $\mathbf{x} \in \mathbb{R}^n$ by $\mathbf{x}^{\mathsf{T}}$. Let $\mathbf{x}^{-i} = (x_1,\ldots, x_{i-1}, x_{i+1}, \ldots,x_n) \in \mathbb{R}^{n-1}$ be the vector including all but the $i^{th}$ element of $\mathbf{x}$. Finally, denote by $\mathds{1}$ the vector of all ones with appropriate size.

\section{The Market Model}\label{sec:model}
We consider a market with a finite number of $M$ consumers and $N$ firms competing for a product. Denote the set of consumers by $\mathcal{M} = \left\lbrace1,2, \ldots,M\right\rbrace$ and the set of suppliers by $\mathcal{N} = \left\lbrace1,2, \ldots,N\right\rbrace$. Let $d_i$ denote consumer $i$'s quantity demanded, which must be greater than a minimum inelastic demand level denoted by $d_0$. Let $s_i$ denote the quantity supplied by firm $i$ that must lie below each supplier's maximum capacity limit denoted by $\kappa_0$. Each consumer receives utility $U_i(d_i)$ for consuming amount $d_i$ and each firm incurs costs $C_i(s_i)$ for producing quantity $s_i$. We make the following assumption on the utility and cost functions.

\textbf{Assumption 1.} For each $i \in \mathcal{M}$, $U_i(d_i)$ is concave, strictly increasing and continuously differentiable for $d_i\geq d_0$ with $U_i(d_0) =0$. For each $i \in \mathcal{N}$, $C_i(s_i)$ is convex, strictly increasing and continuously differentiable with $C_i(s_i)\geq 0$ for $s_i \geq 0$. Over the domain $s_i \leq 0$, $C_i(s_i) = 0$.

The aggregate welfare maximization problem is given by
\begin{subequations}
	\begin{alignat}{2}
	& \underset{\mathbf{d}, \mathbf{s}}{\text{maximize}} && \ \ \mathbb{S}(\mathbf{d}, \mathbf{s}):= ~\sum_{i=1}^{M} U_i(d_i) - \sum_{i=1}^{N} C_{i}(s_{i}) \label{objective}, 
	\\
	& \text{subject to} 
	&& \ \ \sum_{i=1}^{M} d_i = \sum_{i=1}^{N}s_i \label{eq:Clearmarket}, 
	\\
	&&& \ \ 0 \leq s_i \leq \kappa_0, ~\forall~ i=1,\ldots,N, 
	\label{eq:boundary.s}
	\\
	&&& \ \ d_0 \leq d_i, ~\forall~ i=1,\ldots,M, 			
	\label{eq:boundary.d}
	\end{alignat}%
	\label{eq:marketProblem}
\end{subequations}
Henceforth, we will refer to every allocation $\left(\mathbf{d}, \mathbf{s}\right)$ that solves \eqref{eq:marketProblem} as \emph{efficient}. In effect, such allocations can be viewed as those determined by a central entity or market manager that has perfect knowledge on the market and all participants. However, $U_i$ and $C_i$ are generally not available to the market manager.
\begin{figure}
	\centering
	\subfloat[]{\includegraphics[width=0.3\linewidth]{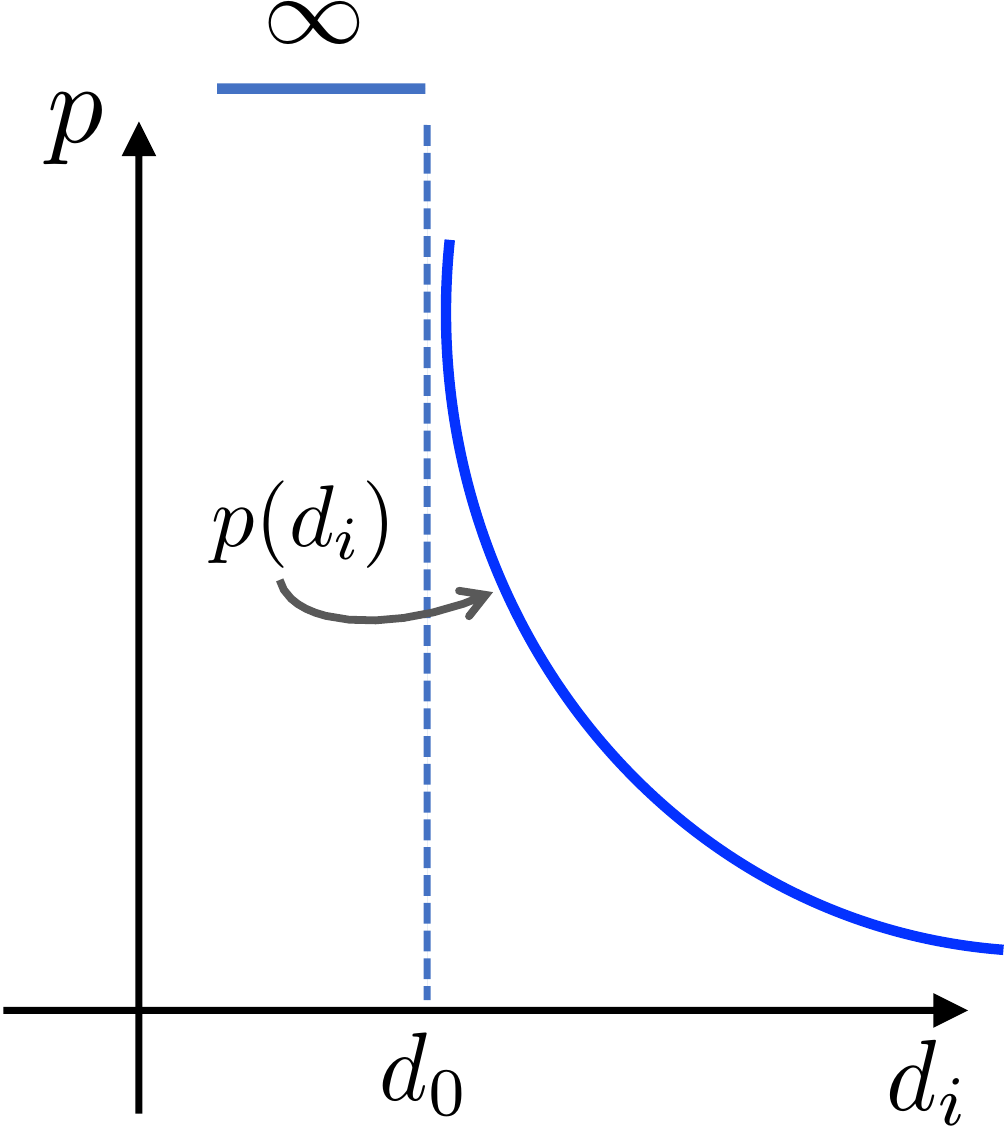} \label{fig:demandBid}}
	\qquad
	\qquad
	\qquad
	\qquad
	\subfloat[]{\includegraphics[width=0.3\linewidth]{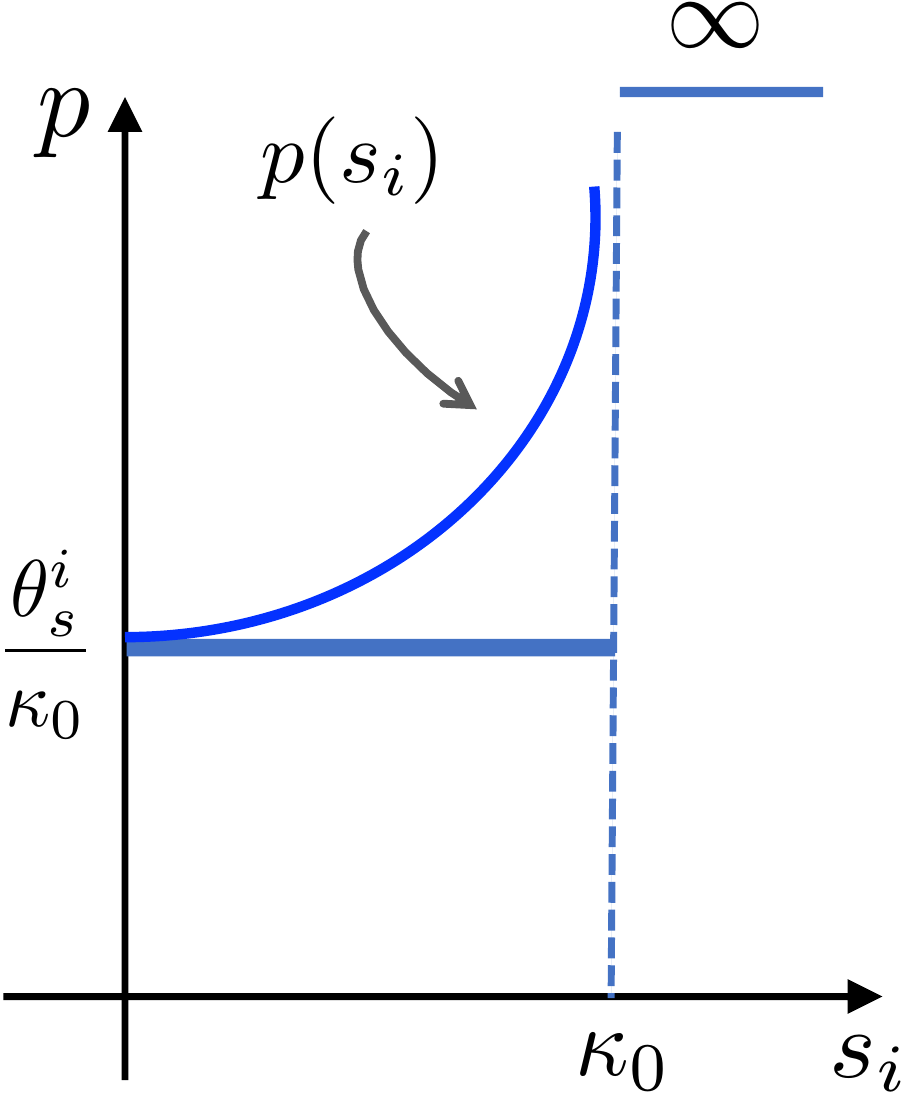} \label{fig:supplyBid}}
	\caption{Illustrations of the demand bid (left) and supply offer (right) structures.}
	\label{fig:Bid}
\end{figure}
Hence, \emph{is there a mechanism that allows market actors to express their preferences in a way that it yields efficient market allocations?}
We consider the following market mechanism for supply and demand allocations. Let consumer $i \in \mathcal{M}$ provide to the market manager a parameter $\theta_d^i \geq 0$. Given price $p>0$, the consumer is willing to buy $d_i=D(\theta_d^i, p)$, where 
\begin{equation}
D(\theta_d^i, p) := d_{0} + \dfrac{\theta_d^i}{p}.
\label{eq:demand.bid}
\end{equation}
The expression in \eqref{eq:demand.bid} represents the quantity the consumer is willing to buy, given the inelastic component $d_0$, the market price $p$, and the parameter $\theta_d^i$. The inelastic demand $d_0$ represents the minimum quantity the consumer must be supplied while ${\theta_d^i}/{p}$ represents the price-responsive portion of her demand. Note that the demand bid is decreasing in price, i.e., it is downward sloping. For ease of exposition we consider equal minimum demand among consumers in the market. The case with distinct $d_0$ is straightforward to generalize. Note that this assumption does not make the consumers homogeneous as each consumer is described by a different utility function.

Let firm $i \in \mathcal{N}$ submit to the market manager a parameter $\theta_s^i \geq 0$. Given price $p>0$, the firm is willing to supply $s_i=S(\theta_s^i, p)$, where  
\begin{equation}
S(\theta_s^i, p) := \kappa_0 - \dfrac{\theta_s^i}{p}.
\label{eq:supply.bid}
\end{equation}
The supply offer \eqref{eq:supply.bid} represents the quantity the firm is willing to supply as a function of price. The supply offer is further parameterized in the capacity $\kappa_0$, which represents the supplier's maximum production capacity. For ease of exposition we consider equal capacities among firms in the market.
We refer to Figure \ref{fig:Bid} for illustrations of how $D(\theta_d^i, p)$ and  $S(\theta_s^i, p)$ vary with price. Observe that as the demand approaches $d_0$, the consumer's willingness to buy approaches infinity. Similarly, as the supply quantity approaches the firm's maximum capacity the requested market price grows large.
%In a sense, the nonlinearity of the supply and demand functions in the price $p$ naturally captures the convexity of cost functions of suppliers and concavity of utility functions of demanders. 

Let $\boldsymbol{\theta}_d =  \left(\theta_d^1, \ldots, \theta_d^M\right)$ and $\boldsymbol{\theta}_s =  \left(\theta_s^1, \ldots, \theta_s^N\right)$ represent the collection of demand bid and supply offer parameters, respectively.
The market manager chooses price $p\left( \boldsymbol{\theta}_d ,\boldsymbol{\theta}_s\right) > 0$ to clear the market such that supply equals demand, i.e.,
\begin{equation}
\sum_{i=1}^{M} D(\theta_d^i, p) = \sum_{i=1}^{N} S(\theta_s^i, p).
\label{eq:balance}
\end{equation}
Such choice is only possible when $\mathds{1}^{\mathsf{T}} \boldsymbol{\theta}_d + \mathds{1}^{\mathsf{T}} \boldsymbol{\theta}_s>0$ in which case the market price is given by
\begin{equation}
p\left( \boldsymbol{\theta}_d ,\boldsymbol{\theta}_s\right) = \dfrac{ \mathds{1}^{\mathsf{T}} \boldsymbol{\theta}_d + \mathds{1}^{\mathsf{T}} \boldsymbol{\theta}_s }{ N\kappa_0 - Md_0 }.
\label{eq:market.price}
\end{equation}
Throughout the paper, we assume $Md_0 < N \kappa_0$ and thus the market price is well-defined. In the case where $\mathds{1}^{\mathsf{T}} \boldsymbol{\theta}_d + \mathds{1}^{\mathsf{T}} \boldsymbol{\theta}_s = 0$, i.e., every market participant submits a zero parameter, we adopt the following conventions
$$
D(0, 0) = d_0 \
\text{and} \
S(0, 0) = \kappa_0.
$$
For markets with a perfectly inelastic demand $D$, the residual supply index ($\RSI$) is often adopted as a suitable indicator of market power. Precisely, the $\RSI$ of firm $i$ measures the capability of the aggregate market capacity---excluding that of $i$---to meet demand $D$. In the model considered here, the inelastic portion of demand is $Md_0$. Mathematically, if
$$\RSI_i := \dfrac{(N-1) \kappa_0}{Md_0}$$ is strictly less than one, then firm $i$ is said to be \emph{pivotal}. See \citet{Newberry} and \citet{swinand} for further details. As we show in Section \ref{sec:Nash}, the presence of pivotal suppliers is critical in the analysis of the market outcome under strategic interactions.

\section{Perfect Competition}\label{sec:pc}
In this section, we study the market outcome assuming all market participants are pure price-takers. We aim to establish the existence and characterization of the competitive market equilibrium taking into account the profit-maximizing nature of market actors. Given market price $\mu>0$ each consumer maximizes the payoff
\begin{equation}
\pi_d^i(\theta_d^i, \mu) = U_i\left(  D\left(\theta_d^i, \mu\right) \right) - \mu D\left(\theta_d^i, \mu\right), ~i\in \mathcal{M}.
\label{eq:demander.payoff}
\end{equation}

Similarly, each supplier maximizes
\begin{equation}
\pi_s^i(\theta_s^i, \mu) = \mu S(\theta_s^i, \mu) - C_{i}(S(\theta_s^i,\mu)), ~i \in \mathcal{N}.
\label{eq:supplier.payoff}
\end{equation}
We now proceed with our first result which shows that when consumers bid in \eqref{eq:demand.bid} and firms offer in \eqref{eq:supply.bid} the market supports an efficient competitive equilibrium.
\begin{theorem}
	\label{prop:CE}
	Suppose Assumption 1 is satisfied. Then, there exists a competitive market equilibrium $\left(\boldsymbol{\theta}^*_d,\boldsymbol{\theta}^*_s,\mu\right)$ satisfying:
	\begin{equation}
	\pi_d^i(\theta_d^{i^*}, \mu) \geq \pi_d^i(\theta_d^i, \mu),~ \forall~ \theta_d^i \geq 0 \text{ and } i \in \mathcal{M}
	\label{eq:EC1}
	\end{equation}
	\begin{equation}
	\pi_s^i(\theta_s^{i^*}, \mu) \geq \pi_s^i(\theta_s^i, \mu),~ \forall~\theta_s^i \geq 0 \text{ and } i \in \mathcal{N}
	\label{eq:EC2}
	\end{equation}
	\begin{equation}
	\mu~\text{is given by (\ref{eq:market.price})}.
	\label{eq:muCondition}
	\end{equation}
	Moreover, the supply vector defined by $s_{i}^* =S(\theta_s^{i^*}, \mu)$  and the demand vector defined by $d_i^*= D\left(\theta_d^{i^*}, \mu\right)$ is an efficient allocation.
\end{theorem}

According to Theorem \ref{prop:CE}, under perfect competition, suppliers and demanders maximize their payoffs and the resulting market allocation is efficient. This implies that given price $\mu$, the firms have no incentive to deviate from supplying $\mathbf{s}^*$ and consumers have no incentive to deviate from buying $\mathbf{d}^*$. Thus the competitive market allocation is efficient and the market clearing price is the shadow value of the constraint $\sum_{i=1}^{M}d_i = \sum_{i=1}^{N}s_i$. In other words, at $\mu$ the marginal social benefit of additional output equals the marginal social cost. The preceding argument establishes the \emph{first fundamental theorem of welfare economics}: if the price $\mu$ and the allocation $\left(\mathbf{d}^*,\mathbf{s}^* \right)$ constitute a competitive equilibrium, then this allocation is efficient.\footnote{In economics the notion of \emph{efficiency} is, typically, equivalent to that of Pareto optimality. In other words, a Pareto optimal market allocation means that there is no other allocation that can make any market actor better off without making another actor worse off.}.

\section{Strategic Suppliers and Demanders}
\label{sec:Nash}
In contrast to the price-taking model, we now consider a model where the market participants are price-anticipating. Price-anticipating suppliers and consumers realize that the market price is a function of their actions and adjust their payoff accordingly. The payoff for the price-anticipating consumer $i \in \mathcal{M}$ is
\begin{equation}
\pi_{d}^i(\theta_{d}^i, \boldsymbol{\theta}_d^{-i}, \boldsymbol{\theta}_s) = U_i\left(d_0 + \dfrac{\theta_d^i}{p\left(\boldsymbol{\theta}_d, \boldsymbol{\theta}_s\right)} \right) - p\left(\boldsymbol{\theta}_d,\boldsymbol{\theta}_s\right) d_0 - \theta_d^i.
\label{eq:demander.Nash.payoff}
\end{equation}
Note that the payoff of each consumer now depends on the actions of all other market participants, that are collectively incorporated in the market price.
Similarly, for each firm the payoff function depends on her action $\theta_s^i$ and the actions of all other market participants. Therefore, the payoff of firm $i \in \mathcal{N}$ is given by
\begin{equation}
\pi_{s}^i(\theta_{s}^i, \boldsymbol{\theta}_s^{-i}, \boldsymbol{\theta}_d) = p\left(\boldsymbol{\theta}_d, \boldsymbol{\theta}_s\right)\kappa_0 - \theta_s^i - C_i\left(\kappa_0 - \dfrac{\theta_s^i}{ p\left(\boldsymbol{\theta}_d, \boldsymbol{\theta}_s\right) }\right).
\label{eq:supplier.Nash.payoff}
\end{equation}
We define the game $\mathcal{G}$ with $\mathcal{M}\cup \mathcal{N}$ denoting the set of \emph{players} with strategy spaces $\boldsymbol{\Theta_i} = \mathbb{R}_+$ and payoffs given by \eqref{eq:demander.Nash.payoff} and \eqref{eq:supplier.Nash.payoff}.
Our goal is to study the existence (and uniqueness) of the Nash equilibria of $\mathcal{G}$ and provide an efficient way to compute the equilibrium. A bid/offer profile $\left(\tilde{\pmb{\theta}}_d, \tilde{\pmb{\theta}}_s\right)$ constitutes a Nash equilibrium if
\begin{align}
\pi_d^i(\tilde{\theta}_d^i, \tilde{\boldsymbol{\theta}}_d^{-i}, \tilde{\boldsymbol{\theta}}_s) \geq \pi_d^i({\theta}_d^i, \tilde{\boldsymbol{\theta}}_d^{-i}, \tilde{\boldsymbol{\theta}}_s),~\forall~\theta_d^i\geq 0 \text{ and } i\in \mathcal{M} \notag \notag\\
\pi_s^i(\tilde{\theta}_s^i, \tilde{\boldsymbol{\theta}}_s^{-i}, \tilde{\boldsymbol{\theta}}_d) \geq \pi_s^i({\theta}_s^i, \tilde{\boldsymbol{\theta}}_s^{-i}, \tilde{\boldsymbol{\theta}}_d),~\forall~ \theta_s^i\geq 0 \text{ and } i\in \mathcal{N}. \notag
\end{align}
We begin with the following result that illustrates how certain market parameters influence the existence of a Nash equilibrium of $\mathcal{G}$.

\begin{lemma} 
	\label{lem:pivotal}
	$\mathcal{G}$ does not admit a Nash equilibrium if a pivotal supplier exists in the market. \label{lem:existence}
\end{lemma}  

In effect, Lemma \ref{lem:pivotal} implies that when $N-1$ firms cannot supply the entire inelastic demand in the market, then there exists a pivotal supplier that faces a non-zero inflexible demand that has infinite willigness to pay. This makes the suppliers' payoff grow unbounded with respect their action $\boldsymbol{\theta}_s$. Hence, a Nash equilibrium cannot exist in this case. As a consequence of Lemma \ref{lem:pivotal}, there cannot exist a Nash equilibrium with $N=1$ since, by definition, the single supplier is pivotal. In view of the above Lemma, we impose the following assumption. 

{\textbf{Assumption 2.} $\RSI_i > 1$ for each firm $i\in \mathcal{N}.$}

Equipped with the previous observations, we present our main result that explicitly characterizes the unique Nash equilibrium of $\mathcal{G}$.

\begin{theorem}
	\label{prop:NE}
	Suppose Assumptions 1-2 hold. $\mathcal{G}$ admits unique Nash equilibrium in $\left(\tilde{\boldsymbol{\theta}}_d, \tilde{\boldsymbol{\theta}}_s\right)$.
	Moreover, the supply profile $$ \tilde{s}_{i} = S_i\left(\tilde{\theta}_s^i, p(\tilde{\boldsymbol{\theta}}_d, \tilde{\boldsymbol{\theta}}_s)\right), i\in\mathcal{N}$$ and the demand profile $$ \tilde{d}_{i} = D_i\left(\tilde{\theta}_d^i, p(\tilde{\boldsymbol{\theta}}_d, \tilde{\boldsymbol{\theta}}_s)\right), i\in\mathcal{M}$$ are given by the unique solution of the following convex program
	\begin{subequations}
		\begin{alignat}{2}
		& \underset{\mathbf{d}, \mathbf{s}}{\text{maximize}} && \ \ \mathbb{\tilde{S}}(\mathbf{d}, \mathbf{s}):= ~\sum_{i=1}^{M} \tilde{U}_i(d_i) - \sum_{i=1}^{N} \tilde{C}_{i}(s_{i}) \label{objective.Nash}, 
		\\
		& \text{subject to} 
		&& \ \ \sum_{i=1}^{M} d_i = \sum_{i=1}^{N}s_i \label{eq:Clearmarket.Nash}, 
		\\
		&&& \ \ 0 \leq s_i \leq \kappa_0, \  i\in\mathcal{N}, 
		\label{eq:boundary.s.Nash}
		\\
		&&& \ \ d_0 \leq d_i, \ i\in\mathcal{M}, 			
		\label{eq:boundary.d.Nash}
		\end{alignat}%
		\label{eq:market.Nash.Problem}
	\end{subequations}
	where 
	\begin{equation}
	\tilde{U}_i(d_i) := \left( 1 - \dfrac{d_i}{N \kappa_0 - (M-1)d_0} \right) U_i(d_i) + \dfrac{1}{N \kappa_0 - (M-1)d_0} \int_{d_0}^{d_i} U_i(z)dz,
	\label{eq:Nash.utility}
	\end{equation}
	\begin{equation}
	\tilde{C}_i(s_i) := \left( 1 + \dfrac{s_i}{(N-1)\kappa_0 - Md_0}\right) C_i(s_i) - \dfrac{1}{(N-1)\kappa_0 - Md_0} \int_{0}^{s_i}C_i(z)dz.
	\label{eq:Nash.cost}
	\end{equation}
\end{theorem}

Computing Nash equilibria is, in general, hard as shown by \cite{Daskalakis}. Theorem \ref{prop:NE} establishes the computation of the market allocation at the Nash equilibrium---and the Nash equilibrium itself---through the solution of a convex program in $\left(\mathbf{d}, \mathbf{s}\right)$ instead of solving M+N problems in the actions $(\boldsymbol{\theta}_d, \boldsymbol{\theta}_s)$, which can be cumbersome depending on the structure of the utility and cost functions. The crux of Theorem \ref{prop:NE} is the construction of an appropriate convex program that yields the market allocation at the Nash equilibrium---a technique closely related to the use of potential functions in characterizing Nash equilibria (\cite{Monderer}). However, the functions \eqref{eq:Nash.utility} and \eqref{eq:Nash.cost} are not potentials for $\mathcal{G}$, since they depend on the allocations and not on the players' decisions. Hence, we cannot use these functions to conclude anything about convergence of best response dynamics to the Nash equilibrium. However, in the following section, we exploit the structure of $\tilde{U}_i$ and $\tilde{C}_i$ to find bounds on the efficiency loss and the markup of prices observed at the Nash equilibrium.

\section{Efficiency Loss And Price Markups}\label{sec:efficiency}

The structure of the modified utility and cost functions allows us to make a number of interesting observations about the behavior of strategic market actors. First, note that since $C_i(s_i)$ are assumed convex and increasing, it follows that $\tilde{C}_i(s_i) \geq C_i(s_i),~\forall~ s_i \geq 0.$ Similarly, since $U_i(d_i)$ are concave and increasing, for each consumer we have $\tilde{U}_i(d_i) \leq U_i(d_i),~\forall~ d_i \geq 0$.
In effect, strategic suppliers misrepresent their costs functions through $\tilde{C}_i(s_i)$, which are greater than the true cost $C_i(s_i)$ at every $s_i$. On the other hand, strategic consumers misrepresent their utilities through $\tilde{U}_i(d_i)$, which are smaller than the true utility $U_i(d_i)$ at every $d_i$.
Moreover, $\mathbb{S}(\tilde{\mathbf{d}}, \tilde{\mathbf{s}}) \leq \mathbb{S}(\mathbf{d}^*, \mathbf{s}^*)$ since the maximum value of $\mathbb{S}$ occurs at $(\mathbf{d}^*, \mathbf{s}^*)$. However, in our next result, we show that the social welfare at the Nash is bounded below and can be relatively close to the optimal value provided some minimum available production capacity.
In order to compute bounds on price markups at the Nash equilibrium we utilize the Lerner index ( see \cite{Le:34}), which we define as
\begin{equation}
\LI(\tilde{\boldsymbol{\theta}}_d, \tilde{\boldsymbol{\theta}}_s) := 1 - \dfrac{1}{p(\tilde{\boldsymbol{\theta}}_d, \tilde{\boldsymbol{\theta}}_s)}\max_i\left\lbrace\dfrac{\partial}{\partial s_i}C_i\left(S(\tilde{\theta}_s^i, p(\tilde{\boldsymbol{\theta}}_d, \tilde{\boldsymbol{\theta}}_s))\right) \right\rbrace.
\label{eq:Lerner}
\end{equation}
The Lerner index measures a firm's market power and it varies from zero to one, with higher values indicating greater market power. The following result summarizes the efficiency loss at the Nash equilibrium and the price markups.

\begin{theorem}   % use the thm environment for theorems
	\label{prop:PoA}
	Suppose Assumptions 1-2  hold. Let $(\mathbf{d}^*, \mathbf{s}^*)$ be the socially optimal allocation from \eqref{eq:marketProblem} and $(\tilde{\mathbf{d}},\tilde{\mathbf{s}})$ be the market allocation at the Nash equilibrium of $\mathcal{G}$. It follows that
	{\small{
			\begin{equation}
			\sum_{i=1}^{M} U_i(\tilde{d}_i) - \sum_{i=1}^{N}C_i(\tilde{s}_i) \geq \dfrac{3}{4}\sum_{i=1}^{M}U_i(d_i^*) - \left(1 - \dfrac{\kappa_0}{\zeta}\right)^{-1}\sum_{i=1}^{N}C_i(s_i^*),\label{eq:social.bound.general}
			\end{equation}}}
	where $\zeta:=N\kappa_0 - Md_0$ and $\zeta \in \left(\kappa_0,\infty\right)$. Moreover, when $\zeta \in \left\lbrack 4 \kappa_0, \infty\right)$ we have
	{\small{
			\begin{equation}
			\sum_{i=1}^{M} U_i(\tilde{d}_i) - \sum_{i=1}^{N}C_i(\tilde{s}_i) \geq \dfrac{3}{4}\sum_{i=1}^{N}U_i(d_i^*) - \dfrac{4}{3}\sum_{i=1}^{N}C_i(s_i^*).\label{eq:social.bound}
			\end{equation}}}
	Finally, the Lerner index at the Nash equilibrium satisfies
	\begin{equation}
	\LI(\tilde{\boldsymbol{\theta}}_d, \tilde{\boldsymbol{\theta}}_s) \leq \dfrac{\kappa_0}{\zeta} < 1. \label{eq:Lerner.bound}
	\end{equation}
\end{theorem}

In effect, Theorem \ref{prop:PoA} provides a lower bound on the social welfare at the Nash equilibrium and an upper bound on the market price with respect to the true marginal cost of suppliers. Notice that $\mathbb{{S}}(\tilde{\mathbf{d}}, \tilde{\mathbf{s}})$ is in the worst case 3/4 of the aggregate utility less $\frac{\zeta}{\zeta-\kappa_0}$ of the aggregate costs at the efficient allocation. We do not claim this bound is tight; there may exist an even tighter bound on the social welfare the computation of which we relegate to future work. Higher values of $\zeta$ yield values of the social welfare at the Nash equilibrium closer to $\mathbb{{S}}({\mathbf{d}}^*,{\mathbf{s}}^*)$. The worst-case values for $\mathbb{S}(\tilde{\mathbf{d}}, \tilde{\mathbf{s}})$ arise when $\zeta \rightarrow \kappa_0$, although it never reaches it. Intuitively, when the aggregate production capacity of supply is relatively close to the total inelastic demand, then firms' market power increases over consumers, gradually inducing \emph{pivotalness} as $\zeta \rightarrow \kappa_0$.
Specifically, for $\zeta \in \left(\kappa_0, 2\kappa_0\right)$ the efficiency loss can be arbitrarily high, similar to that derived by \cite{Xu} for a market with capacity-constrained suppliers. When $\zeta \in \left\lbrack2\kappa_0, \infty\right)$ the worst-case aggregate cost coefficient in \eqref{eq:social.bound.general} is equal to two and we recover the worst-case bound of \cite{Johari:11} derived for uncapacitated supply function competition.
Moreover, \eqref{eq:social.bound} shows that provided some minimum available production capacity, the social welfare at the Nash equilibrium is no lower than 3/4 of the aggregate utility less 4/3 of the aggregate cost at the efficient allocation, which is not much lower than $\mathbb{{S}}({\mathbf{d}}^*,{\mathbf{s}}^*)$. 
From \eqref{eq:Lerner.bound} note that the Lerner index is strictly less than one due to the non-pivotal supplier assumption. As $\zeta$ grows large, $\LI(\tilde{\boldsymbol{\theta}}_d, \tilde{\boldsymbol{\theta}}_s)$ goes to zero, indicating less market power on the supply side. As $Md_0$ approaches $N\kappa_0$, the index grows large implying high market power since there is little available capacity to supply anything more than the total inelastic demand.

%------------------------------

\section{Illustrative Examples}
\label{sec:numerics}
In this section we provide numerical experiments to illustrate the behavior of the social welfare under perfect competition and strategic interactions with respect to specific problem parameters. As shown in Section \ref{sec:efficiency}, the key parameter that affects social welfare is the total flexible capacity in the market $\zeta$.

Consider a market with $N=6$ and $M=5$. Let each consumer $i\in\mathcal{M}$ have utility 
$$U_i(d_i) := \beta_i\log(d_i), \qquad d_0 = 1.$$ Note that the above utility function is strictly concave and increasing and attains a minimum value $U_i(d_0) = 0$ for every $i \in \mathcal{M}$. Moreover, every supplier $j \in \mathcal{N}$ incurs costs given by $$C_j(s_j) := \frac{1}{2} \alpha_j s_j^2.$$

The modified utility for each $i \in \mathcal{M}$ at which the Nash equilibrium can be computed via \eqref{eq:market.Nash.Problem} is
\begin{align*} \tilde{U}_i(d_i) &= \left(1 - \dfrac{d_i}{\zeta + d_0}\right) \beta_i\log(d_i) \\
&\qquad\qquad +\dfrac{\beta_i}{\zeta+d_0}\left( d_i\log(d_i) - d_i + 1 \right)
\end{align*} Similarly, for each $j \in \mathcal{N}$ the modified cost is given by
\begin{align*} &\tilde{C}_j(s_j) = \left(1 + \dfrac{s_j}{\zeta - \kappa_0}\right)\dfrac{1}{2}\alpha_js_j^2- \frac{1}{6}\left(\dfrac{\alpha_j}{\zeta - \kappa_0}\right)s_j^3.
\end{align*}

\begin{figure*}
	\centering
	\subfloat[]{\includegraphics[width=0.3\linewidth]{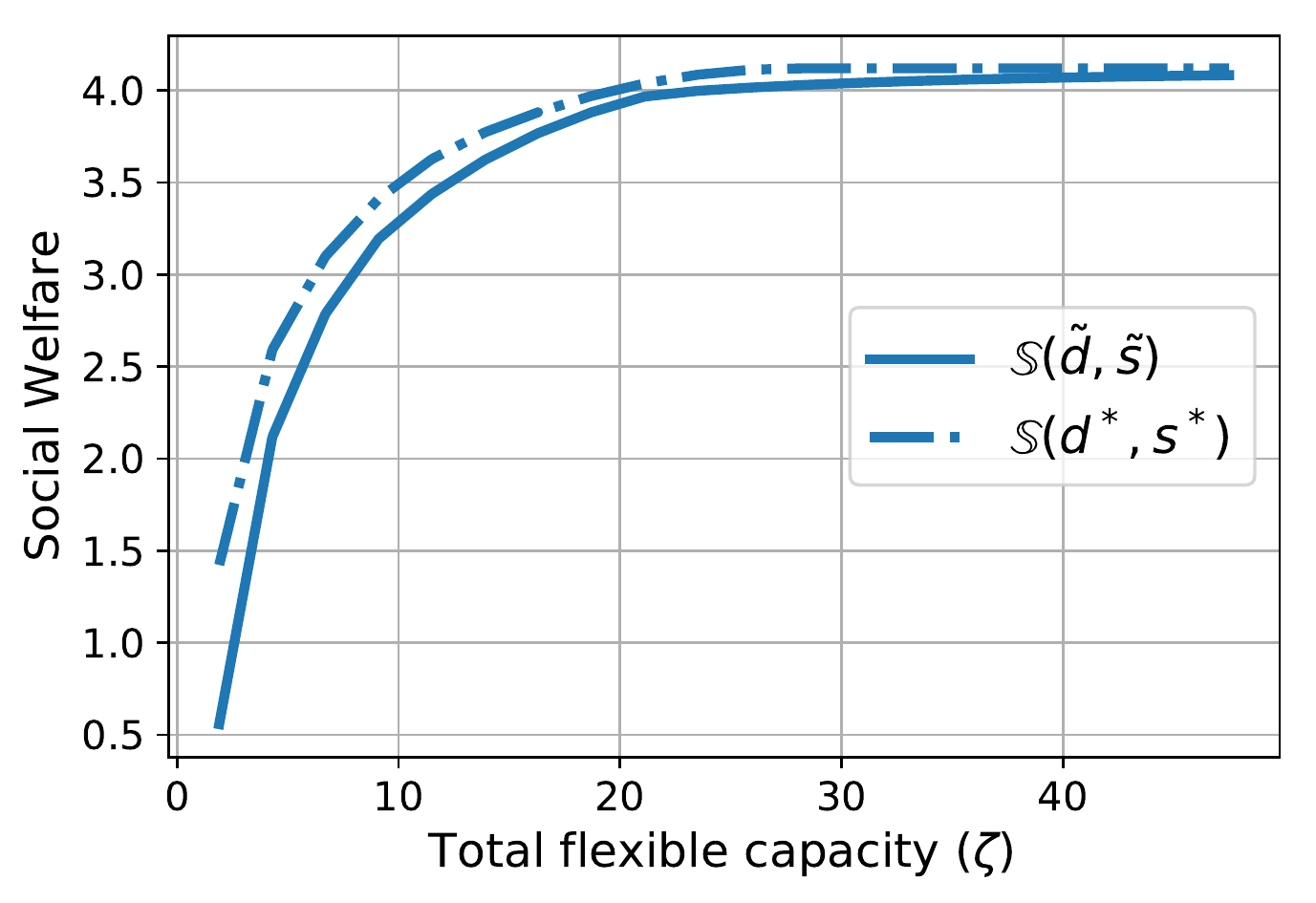} \label{fig:swNash}}
	\subfloat[]{\includegraphics[width=0.3\linewidth]{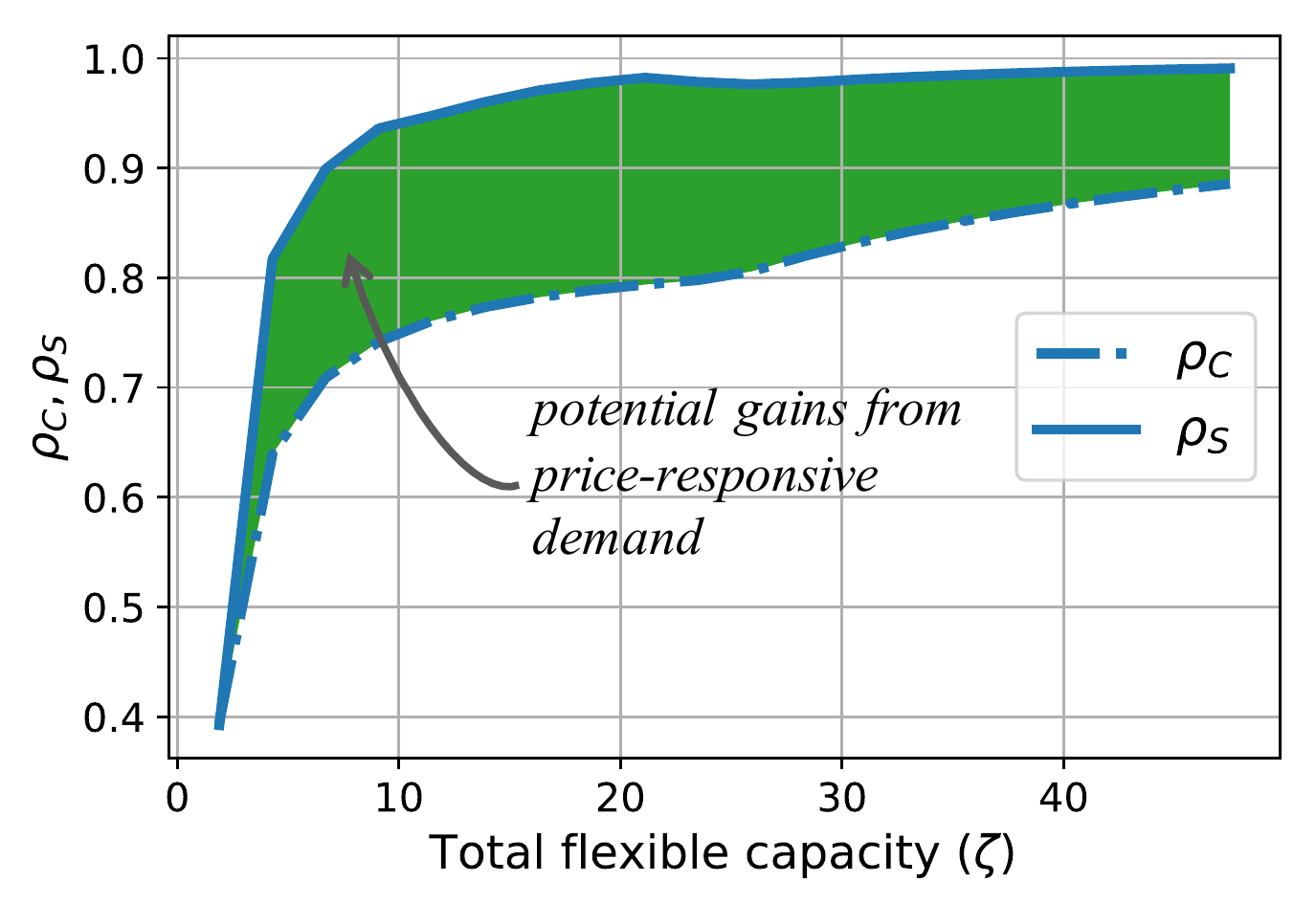} \label{fig:PoA}}
	\subfloat[]{\includegraphics[width=0.3\linewidth]{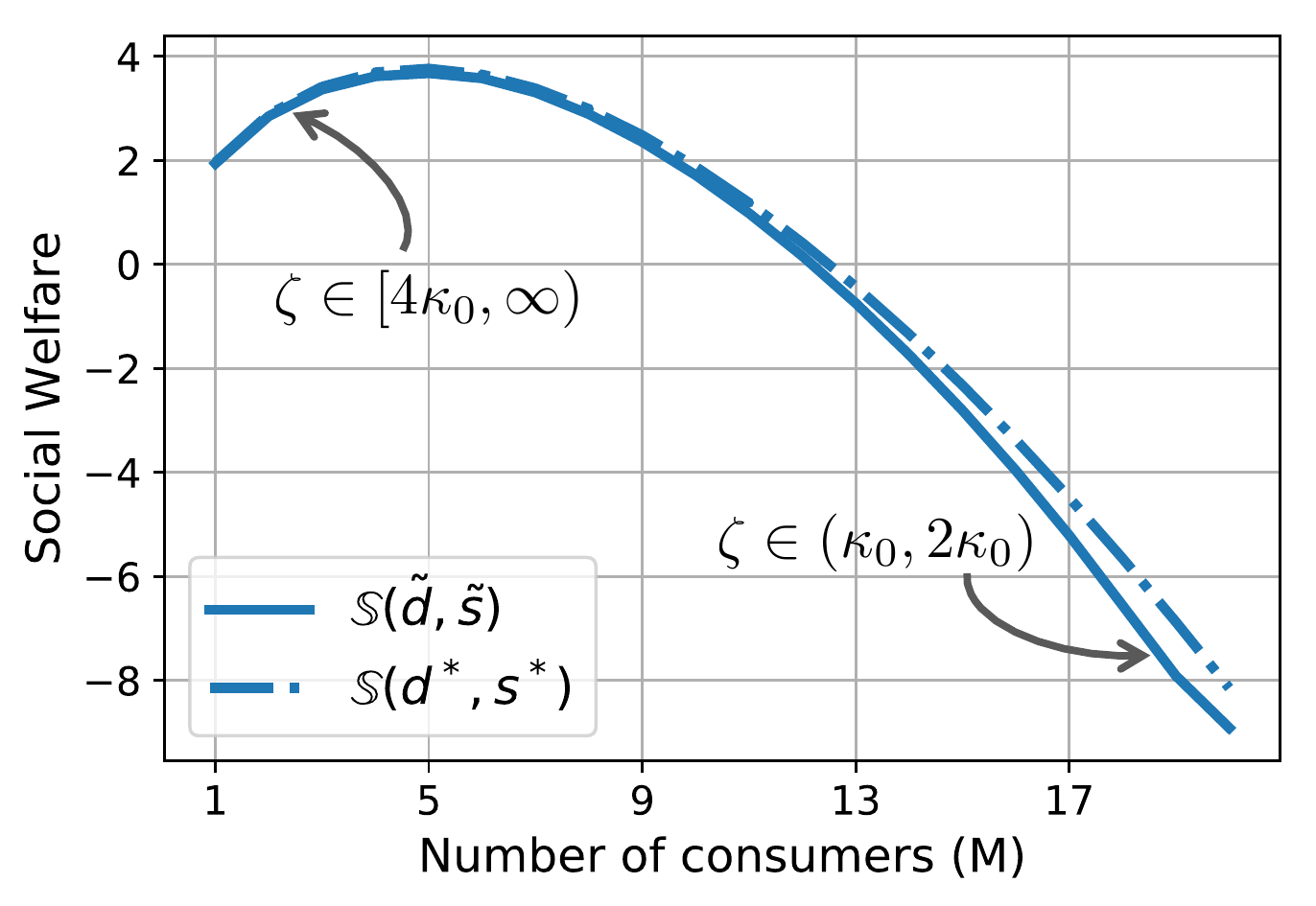} \label{fig:swNash_2}}
	\caption{Plot (a) shows values of the social welfare with respect to $\zeta$ at the efficient and Nash equilibrium allocations. In (b) we plot social welfare bounds for (strategic) price-responsive and perfectly inelastic demand. Plot (c) shows how the social welfare varies with respect to the number of consumers.}
	\label{fig:wholefiglabel}
\end{figure*}

Figures \ref{fig:swNash} and \ref{fig:PoA} illustrate how social welfare under perfect competition and at the Nash equilibrium varies with respect to $\zeta$.\footnote{For the experiments we assumed that the vector of utility coefficients $\beta_i$ is $\left\lbrack 1, 1, 1.5, 2, 2\right\rbrack$ and of the cost coefficients $\alpha_j$ is $\left\lbrack 0.1, 0.2, 0.3, 0.4, 0.5, 0.5 \right\rbrack$.} More specifically, we start with a value of $\kappa_0 = 1.1$---just slightly higher than $d_0$ and to avoid pivotal suppliers---and increase it gradually. Observe that the higher the value of $\zeta$ the closer $\mathbb{S}(\tilde{\mathbf{d}}, \tilde{\mathbf{s}})$ is to $\mathbb{S}(\mathbf{d}^*, \mathbf{s}^*)$.  On the other hand, the smaller $\zeta$ is, the higher the efficiency loss at the Nash equilibrium. To gain additional insights, define the following ratio $$\rho_{S} := \frac{\mathbb{S}(\tilde{\mathbf{d}},\tilde{\mathbf{s}})}{\mathbb{S}(\mathbf{d}^*,\mathbf{s}^*)}.$$
For the special case in which the market has perfectly inelastic, non-strategic demand, we utilize the wost-case market performance metric $\rho_{C}$, which is adjusted from \cite{Xu} and is given by
\begin{equation}
\rho_{C} = \left(1 + \dfrac{1}{\zeta-\kappa_0}\min\left\lbrace\kappa_0, Md_0\right\rbrace\right)^{-1}.
\end{equation}

Figure \ref{fig:PoA} demonstrates that the worst-case value of $\mathbb{S}(\tilde{\mathbf{d}},\tilde{\mathbf{s}})$ occurs when $\zeta = 1.6 \in \left(\kappa_0, 2\kappa_0\right)$ where the ratio $\rho_{S} =0.4$. Immediately after $\zeta \in \left\lbrack2\kappa_0, \infty\right)$, the ratio $\rho_{S}$ jumps to $0.8$ and stays above $0.9$ after $\zeta \geq 4 \kappa_0$. 

Note that $\rho_{S}$ lies everywhere above $\rho_{C}$ except when $\zeta \in (\kappa_0, 2\kappa_0)$ where $\rho_{S}$ =$\rho_{C}$. This implies that although consumers are strategic, the market efficiency loss is lower-bounded by the worst-case performance of a market with perfectly inelastic demand. It remains to be shown whether this outcome holds more broadly, for any choice of cost and utility functions.
Finally, increasing the number of consumers, while keeping the production capacity constant, widens the disparity between $\mathbb{S}(\tilde{\mathbf{d}},\tilde{\mathbf{s}})$ and $\mathbb{S}({\mathbf{d}}^*,{\mathbf{s}}^*)$ as shown in Figure \ref{fig:swNash_2}. This illustrates the effect of increasing the inelastic portion of demand and as such inducing higher market power on the existing set of firms, which is also captured by the Lerner index in \eqref{eq:Lerner.bound}.

\section{Conclusion}\label{sec:conclusion}
We studied a market with $N$ suppliers and $M$ consumers that compete in supply offers and demand bids for a product. Our analysis showed that with a specific family of scalar-parameterized offers/bids, the market supports an efficient competitive equilibrium. Under strategic interactions, we showed there exists a unique Nash equilibrium and propose an efficient way of computing the induced market allocation. Moreover, the welfare loss and the price markups at the Nash equilibrium are bounded.
Understanding how uncertainty on the supply capacity and minimum demand affects the market outcome is an interesting direction for future research. Furthermore, a study of the market competition and efficiency loss when only one side (demand or supply) is strategic would complete the analysis of the deterministic model. The market mechanism presented here has multiple interesting applications. For example, owing to their simplicity, scalar-parameterized offers/bids can be effectively utilized to model competition among retail electricity customers that are becoming both consumers and producers, due to the proliferation of distributed energy resources.

%\section{Acknowledgements}

\bibliographystyle{abbrvnat}
\bibliography{myrefs}

%----------------------------

\appendix

\section{Proof of Theorem \ref{prop:CE}.}
\label{sec:proofMain}
The crux of our derivations relies on Lagrangian duality to establish that the equilibrium conditions of \eqref{eq:EC1}-\eqref{eq:EC2} together with \eqref{eq:muCondition} are equivalent to the optimality conditions of \eqref{eq:marketProblem}.
We begin with the consumer's problem. The payoff in \eqref{eq:demander.payoff} is concave in each player's action $\theta_d^i$. 
Hence, the Karush-Kuhn-Tucker (KKT) optimality conditions are both necessary and sufficient. An optimal strategy $\theta_d^{i^*} \geq 0$ must satisfy
\begin{subequations}
	\begin{equation}
	\dfrac{\partial U_i\left(D(\theta_d^{i^*}, \mu)\right)}{\partial d_i} = \mu,~~\text{if } \theta_d^{i^*}>0 
	\end{equation}
	\begin{equation}
	\dfrac{\partial U_i\left(D(\theta_d^{i^*}, \mu)\right)}{\partial d_i} \leq \mu,~~\text{if } \theta_d^{i^*}=0
	\end{equation}
	\label{eq:demander.KKT}
\end{subequations}
Each supplier's payoff is concave in the action $\theta_s^i$. Moreover, an optimal strategy $\theta_s^{i^*}$ must lie in the closed interval $\left[0, \mu \kappa_0\right]$. If not, then it is easy to show that $S_i(\theta_s^i, \mu) <0$ for $\theta_s^i > \mu \kappa_0$. Therefore, such strategies cannot occur at the equilibrium since they yield negative payoff. As such, an optimal strategy $\theta_s^{i^*}$ must satisfy
\begin{subequations}
	\begin{equation}
	\dfrac{\partial C_i\left(S(\theta_s^{i^*}, \mu)\right)}{\partial s_i} \leq \mu,~~\text{if } 0 \leq \theta_s^{i^*} < \mu \kappa_0
	\end{equation}
	\begin{equation}
	\dfrac{\partial C_i\left(S(\theta_s^{i^*}, \mu)\right)}{\partial s_i} \geq \mu,~~\text{if } 0 < \theta_s^{i^*} \leq \mu \kappa_0
	\end{equation}
	\label{eq:supplier.KKT}
\end{subequations}
We now turn to problem \eqref{eq:marketProblem} solved by the market manager. Associate Lagrange multiplier $\lambda$ with the equality constraint \eqref{eq:Clearmarket}. The objective function is continuous and concave over a compact convex set. Therefore, there exists at least one optimal solution $\left(\mathbf{d}^*, \mathbf{s}^*\right)$ and $\lambda^* \geq 0$ that satisfy
\begin{subequations}
	\begin{equation}
	\dfrac{\partial U_i(d^*_i)}{\partial d_i} = \lambda^*,~~\text{if } d_i^*>d_0 
	\end{equation}
	\begin{equation}
	\dfrac{\partial U_i(d_i^*)}{\partial d_i} \leq \lambda^*,~~\text{if } d_i^*=d_0
	\end{equation}
	\label{eq:manager.D.KKT}
\end{subequations}
Similarly, the supply vector $\mathbf{s}^*$ must satisfy
\begin{subequations}
	\begin{equation}
	\dfrac{\partial C_i(s^*_i)}{\partial s_i} \geq \lambda^*,~~\text{if } 0 \leq s_i^* < \kappa_0
	\end{equation}
	\begin{equation}
	\dfrac{\partial C_i(s_i^*)}{\partial s_i} \leq \lambda^*,~~\text{if } 0 < s_i^* \leq \kappa_0
	\end{equation}
	\label{eq:manager.S.KKT}
\end{subequations}
Primal feasibility requires
\begin{equation}
\sum_{i=1}^{M} d_i^* = \sum_{i=1}^{N}s_i^*.
\label{eq:manager.primal}
\end{equation}
Note that $\lambda^* > 0$ since $U_i$ and $C_i$ are strictly increasing and there exists at least one $s_i^*>0$.
If the pair $\left(\mathbf{s}^*, \lambda^*\right)$ satisfies \eqref{eq:manager.S.KKT} and we let $\theta_s^i = \lambda^*\left(\kappa_0 - s_i^*\right)$ then $(\boldsymbol{\theta}_s, \lambda^*)$ satisfy \eqref{eq:supplier.KKT} and $\boldsymbol{\theta}_s \geq 0$. In effect \eqref{eq:manager.S.KKT} become equivalent to \eqref{eq:supplier.KKT}.
Similarly, if the pair $\left( \mathbf{d}^*,\lambda^* \right)$ satisifes \eqref{eq:manager.D.KKT} and we let $\theta_d^i = \lambda^* (d_i^*- d_0)$ then $(\boldsymbol{\theta}_d, \lambda^*)$ satisfy \eqref{eq:demander.KKT} and $\boldsymbol{\theta}_d \geq 0$. In this case, \eqref{eq:manager.D.KKT} become equivalent to \eqref{eq:demander.KKT}. Finally, the market clearing condition in \eqref{eq:manager.primal} yields $\lambda^* = \mu$. Hence, $(\boldsymbol{\theta}_d,\boldsymbol{\theta}_s,\mu)$ is a competitive equilibrium.
Now suppose that $(\boldsymbol{\theta}^*_d, \boldsymbol{\theta}^*_s, \mu)$ satisfy \eqref{eq:demander.KKT},\eqref{eq:supplier.KKT} and \eqref{eq:market.price}. Let $s_i = S(\mu, \theta_s^{i^*})$ for $i \in \mathcal{N}$ and $d_i = D(\mu, \theta_d^{i^*})$ for $i \in \mathcal{M}$. Then, it is easy to verify that the vector $(\mathbf{d}, \mathbf{s})$ satisfies \eqref{eq:manager.D.KKT} and \eqref{eq:manager.S.KKT}. Therefore, $(\mathbf{d}, \mathbf{s})$ is an efficient allocation.

\section{Proof of Lemma \ref{lem:pivotal}.}
Let firm $i$ be a pivotal supplier. Then it must hold
\begin{equation}
\RSI_i = \dfrac{(N-1) \kappa_0}{M d_0} < 1,
\end{equation}
i.e., the total production capacity less that of $i$'s is less than the total inelastic demand in the market. In this case, the first derivative of the supplier's payoff becomes
\begin{equation}
\dfrac{\partial \pi_s^i(\theta_{s}^i, \boldsymbol{\theta}_s^{-i}, \boldsymbol{\theta}_d)}{\partial \theta_s^i} = \dfrac{\kappa_0}{N\kappa_0 - Md_0} - 1 + (N\kappa_0 - Md_0)\dfrac{\partial C_i}{\partial s_i}\left( \dfrac{ \mathds{1}^{\mathsf{T}} \boldsymbol{\theta}_d + \mathds{1}^{\mathsf{T}} \boldsymbol{\theta}_s^{-i}}{(\mathds{1}^{\mathsf{T}} \boldsymbol{\theta}_d + \mathds{1}^{\mathsf{T}} \boldsymbol{\theta}_s)^2}\right)>0.
\end{equation}
Therefore, the payoff is strictly increasing in the action $\theta_s^i$ and grows unbounded. A Nash equilibrium does not exist.	

\section{Proof of Theorem \ref{prop:NE}.}
We break the proof into five steps. First, we show that any Nash equilibrium has at least two positive components and we derive the necessary and sufficient conditions for such equilibrium. Next we establish the existence and uniqueness of the market allocation at the Nash equilibrium and derive the optimality conditions for \eqref{eq:market.Nash.Problem}. We show that under the demand bid \eqref{eq:demand.bid} and the supply offer \eqref{eq:supply.bid} the equilibrium conditions of all players become equivalent to the optimality conditions of \eqref{eq:market.Nash.Problem}. Finally, we establish uniqueness of the Nash equilibrium. 

{\bf Step 1. Any Nash Equilibrium Has at Least Two Positive Components.}

First, it is straightforward to see that $\mathds{1}^{\mathsf{T}} \boldsymbol{\theta}_d + \mathds{1}^{\mathsf{T}} \boldsymbol{\theta}_s = 0$ cannot occur at the Nash equilibrium since $N\kappa_0 > Md_0$ and therefore the market does not clear.
Next, we consider two cases. First, assume that $\mathds{1}^{\mathsf{T}} \boldsymbol{\theta}_d = 0$. Fix firm $j$ and let $\mathds{1}^{\mathsf{T}} \boldsymbol{\theta}_s^{-j} = 0$. Note that, in this case, $\theta_s^j > 0$ is not possible by the non-pivotal supplier assumption. A Nash equilibrium cannot exist with all consumers bidding zero and all but one supplier offering a strictly positive $\theta_s^i$.
Second, assume $\mathds{1}^{\mathsf{T}} \boldsymbol{\theta}_s = 0$. Fix consumer $j$ and let $\mathds{1}^{\mathsf{T}} \boldsymbol{\theta}_d^{-j} = 0$. Then, $\theta_d^j >0$ implies
$d_j > d_0$. Hence, consumer $j$ faces a total available residual supply equal to $N\kappa_0 - Md_0$. In this case, the payoff of consumer $j$ is given by
\begin{equation}
U_j\left(d_0 + N\kappa_0 - Md_0\right) - \dfrac{d_0}{N\kappa_0 - Md_0}\theta_d^j - \theta_d^j,
\end{equation}
which is strictly increasing as $\theta_d^j$ becomes small and attains its maximum when $\theta_d^j =0$. Thus for any $\theta_d^j>0$ there exists an infinitesimally smaller and positive $\theta_d^j$ that yields higher payoff. Moreover, by definition $U_j(0, 0) = U(d_0) = 0$. A Nash equilibrium does not exist in this case. Hence, at the Nash equilibrium, the vector $\boldsymbol{\theta} = \left(\boldsymbol{\theta}_d, \boldsymbol{\theta}_s\right)$ has at least two positive components.

{\bf Step 2. Necessary and Sufficient Nash Equilibrium Conditions.}
Having shown that any Nash equilibrium must have at least two positive components, we only focus in the region where $\mathds{1}^{\mathsf{T}} \boldsymbol{\theta}_d + \mathds{1}^{\mathsf{T}} \boldsymbol{\theta}_s > 0$. Note that, for each consumer (firm), her payoff is strictly concave in the action $\theta_d^i$ ($\theta_s^i$).
Hence, the KKT optimality conditions are both necessary and sufficient. Moreover, we must have $$0 \leq \tilde{\theta}_s^i \leq \theta_{\text{max}}^i:=\dfrac{\kappa_0}{(N-1)\kappa_0 - Md_0} \left(\sum_{i=1}^{M}\theta_d^i + \sum_{j\neq i}^{N} \theta_s^j\right),$$ in order for $S\left( \tilde{\theta}_s^i, p(\boldsymbol{\theta}) \right) \geq 0$. 
We have the following equilibrium conditions.

A demand profile $\tilde{\boldsymbol{\theta}}_d = \left(\tilde{\theta}_d^1,\ldots, \tilde{\theta}_d^N\right)$ is a Nash profile if and only if
{\small{
		\begin{subequations}
			\begin{equation}
			\left( 1- \dfrac{D(\tilde{\theta}_d^i, p(\tilde{\boldsymbol{\theta}}_d,\tilde{\boldsymbol{\theta}}_s))}{N\kappa_0 - (M-1)d_0} \right)\dfrac{\partial U_i\left(D(\tilde{\theta}_{d}^i, p(\tilde{\boldsymbol{\theta}}_d,\tilde{\boldsymbol{\theta}}_s))\right)}{\partial d_i} = p(\tilde{\boldsymbol{\theta}}_d,\tilde{\boldsymbol{\theta}}_s),\text{if } \tilde{\theta}_{d}^i>0 
			\end{equation}
			\begin{equation}
			\left( 1- \dfrac{D(\tilde{\theta}_d^i, p(\tilde{\boldsymbol{\theta}}_d,\tilde{\boldsymbol{\theta}}_s))}{N\kappa_0 - (M-1)d_0} \right)\dfrac{\partial U_i\left(D(\tilde{\theta}_{d}^i,p(\tilde{\boldsymbol{\theta}}_d,\tilde{\boldsymbol{\theta}}_s) )\right)}{\partial d_i} \leq p(\tilde{\boldsymbol{\theta}}_d,\tilde{\boldsymbol{\theta}}_s),\text{if }\tilde{\theta}_{d}^i=0
			\end{equation}
			\label{eq:demander.Nash.KKT}
		\end{subequations}}}
		%The above equilibrium conditions resemble those of a price-taking consumer with some modifed marginal utility. Essentially we can interpret \eqref{eq:demander.Nash.KKT} as if the consumer is willing to purchase an non-zero quantity whenever her modified marginal utility is equal to the market price.
		A supply profile $\tilde{\boldsymbol{\theta}}_s^i = \left(\tilde{\theta}_s^1,\ldots, \tilde{\theta}_s^N\right)$ is a Nash equilibrium if and only if
		{\small{
				\begin{subequations}
					\begin{equation}
					\left( 1 + \dfrac{S\left( \tilde{\theta}_s^i, p(\tilde{\boldsymbol{\theta}}_d,\tilde{\boldsymbol{\theta}}_s) \right)}{(N-1)\kappa_0 - Md_0} \right) \dfrac{\partial C_i\left(S(\tilde{\theta}_{s}^i, p(\tilde{\boldsymbol{\theta}}_d,\tilde{\boldsymbol{\theta}}_s)))\right)}{\partial s_i} \leq p(\tilde{\boldsymbol{\theta}}_d,\tilde{\boldsymbol{\theta}}_s), \text{if } 0 \leq \tilde{\theta}_{s}^i< \theta_{\text{max}}^i
					\end{equation}
					\begin{equation}
					\left( 1 + \dfrac{S\left( \tilde{\theta}_s^i, p(\tilde{\boldsymbol{\theta}}_d,\tilde{\boldsymbol{\theta}}_s)) \right)}{(N-1)\kappa_0 - Md_0} \right) \dfrac{\partial C_i\left(S(\tilde{\theta}_{s}^i, p(\tilde{\boldsymbol{\theta}}_d,\tilde{\boldsymbol{\theta}}_s)))\right)}{\partial s_i} \geq p(\tilde{\boldsymbol{\theta}}_d,\tilde{\boldsymbol{\theta}}_s), \text{if } 0 < \tilde{\theta}_{s}^i \leq \theta_{\text{max}}^i
					\end{equation}
					\label{eq:supplier.Nash.KKT}
				\end{subequations}}}
				The equilibrium conditions \eqref{eq:demander.Nash.KKT} and \eqref{eq:supplier.Nash.KKT} are derived from the KKT conditions of each player's payoff maximization problem, where the payoff of each consumer and supplier is given by expressions \eqref{eq:demander.Nash.payoff} and \eqref{eq:supplier.Nash.payoff} respectively.
 
%------------------

{\bf Step 3. Existence and Uniqueness of the Nash Market Allocation.}
Equipped with the above relations we now proceed to the market manager's problem. Note that $\tilde{U}_i(d_i)$ is strictly concave and $\tilde{C}_i(s_i)$ is strictly convex. Hence, the objective function $\widetilde{\mathbb{S}}(\mathbf{d}, \mathbf{s})$ is continuous and strictly concave over a compact set. 
Specifically, the Hessian matrix $\mathbf{H}$ of $\tilde{\mathbb{S}}(\mathbf{d}, \mathbf{s})$ has diagonal elements
\begin{equation}
h_{ii} = \begin{cases}
\dfrac{\partial^2 \tilde{U}_i(d_i)}{\partial d_i^2}<0 ,~\text{for } i = 1,\ldots,M\\
-\dfrac{\partial^2 \tilde{C}_i(s_i)}{\partial s_i^2}<0,~\text{for } i=M+1,\ldots,M+N
\end{cases}
\end{equation}
and $h_{ij} = 0$ for $i\neq j$. Hence, $\mathbf{H}$ is negative definite and there exists a unique optimal solution $(\tilde{\mathbf{d}}, \tilde{\mathbf{s}})$ to \eqref{eq:market.Nash.Problem}.

{\bf Step 4. Necessary and Sufficient Conditions for the Market Allocation.}
Let $(\tilde{\mathbf{d}}, \tilde{\mathbf{s}})$ be the unique optimal solution to \eqref{eq:market.Nash.Problem}. There exists a Lagrange mutliplier $\tilde{\lambda}$ such that
{\small{
		\begin{subequations}
			\begin{equation}
			\left( 1- \dfrac{\tilde{d_i}}{N\kappa_0-(M-1)d_0} \right)\dfrac{\partial U_i(\tilde{d_i})}{\partial d_i} = \tilde{\lambda},\text{ if } \tilde{d_i}>d_0 \label{eq:KKT.Nash.1}
			\end{equation}
			\begin{equation}
			\left( 1- \dfrac{\tilde{d_i}}{N\kappa_0 - (M-1)d_0} \right)\dfrac{\partial U_i(\tilde{d_i})}{\partial d_i} \leq \tilde{\lambda},\text{ if }\tilde{d_i}=d_0 \label{eq:KKT.Nash.2}
			\end{equation}
			\begin{equation}
			\left(1+\dfrac{\tilde{s_i}}{(N-1)\kappa_0-Md_0} \right)\dfrac{\partial C_i(\tilde{s_i})}{\partial s_i} \geq \tilde{\lambda},\text{ if } 0 \leq \tilde{s_i} <\kappa_0
			\end{equation}
			\begin{equation}
			\left(1+\dfrac{\tilde{s_i}}{(N-1)\kappa_0-Md_0} \right)\dfrac{\partial C_i(\tilde{s_i})}{\partial s_i} \leq \tilde{\lambda},\text{ if } 0 < \tilde{s_i} \leq \kappa_0
			\end{equation}
			\begin{equation}
			\sum_{i=1}^{M}\tilde{d_i} = \sum_{i=1}^{N} \tilde{s_i}.
			\end{equation}
			\label{eq:manager.Nash.KKT}
		\end{subequations}}}

Note that since there is at least one $\tilde{s}_i>0$ and $U_i$ and $C_i$ are strictly increasing, then $\tilde{\lambda} >0$. Consider the action vectors $\tilde{\theta}_d^i = \tilde{\lambda}(\tilde{d}_i - d_0)$ for $i \in \mathcal{M}$ and $\tilde{\theta}_s^i = \tilde{\lambda}(\kappa_0 - \tilde{s}_i)$ for $i\in \mathcal{N}$. Note that $\theta_d^i \geq 0$ and $\theta_s^i \geq 0$ for every consumer and every firm respectively. 
Suppose now that $d_i > d_0$ and $d_j = d_0$ for $j\neq i$ and let $s_i = \kappa_0$ for $i \in \mathcal{N}$. This implies that $d_i = d_0 + N\kappa_0 - Md_0$. Then from \eqref{eq:KKT.Nash.1} we have $\tilde{\lambda} = 0$. However, we have $\dfrac{\partial U_j(d_0)}{\partial d_j} > 0$ for each $j\in\mathcal{M}$. Therefore, \eqref{eq:KKT.Nash.2} cannot hold for every $j \neq i$. Thus, the vector $\left(\tilde{\boldsymbol{\theta}}_d, \tilde{\boldsymbol{\theta}}_s\right)$ cannot have all components zero except one $\theta_d^i>0$. Similarly, $\left(\tilde{\boldsymbol{\theta}}_d, \tilde{\boldsymbol{\theta}}_s\right)$ cannot have all components zero except one $\theta_s^i >0$ for some firm $i \in \mathcal{N}$. This is obvious by the non-pivotal supplier assumption since it holds $(N-1)\kappa_0 >Md_0$ for every supplier $i$. Hence, at least two components of $\left(\tilde{\boldsymbol{\theta}}_d, \tilde{\boldsymbol{\theta}}_s\right)$ are positive. Moreover, since $\tilde{s}_i = \kappa_0$ if and only if $\theta_s^i =0$, $\tilde{s}_i = 0$ if and only if $\theta_s^i =\theta_{\text{max}}^i$, then it is not hard to see that \eqref{eq:manager.Nash.KKT} become equivalent to \eqref{eq:demander.Nash.KKT}-\eqref{eq:supplier.Nash.KKT}. Hence, the action vector $(\tilde{\boldsymbol{\theta}}_d, \tilde{\boldsymbol{\theta}}_s)$ is a Nash equilibrium. This also establishes existence of the Nash equilibrium.

We now reverse the argument. Let $(\tilde{\boldsymbol{\theta}}_d, \tilde{\boldsymbol{\theta}}_s)$ be a Nash equilibrium profile. That is, it satisfies \eqref{eq:demander.Nash.KKT}-\eqref{eq:supplier.Nash.KKT}. Therefore, it has at least two positive components and $p(\tilde{\boldsymbol{\theta}}_d, \tilde{\boldsymbol{\theta}}_s)>0$. Define the demand allocation $\tilde{d}_i = d_0 + \dfrac{\tilde{\theta}_d^i}{p(\tilde{\boldsymbol{\theta}}_d, \tilde{\boldsymbol{\theta}}_s)}$ for $i\in \mathcal{M}$ and the supply allocation $\tilde{s}_i = \kappa_0 - \dfrac{\tilde{\theta}_s^i}{p(\tilde{\boldsymbol{\theta}}_d, \tilde{\boldsymbol{\theta}}_s)}$ for $i \in \mathcal{N}$. It follows that $(\tilde{\mathbf{d}}, \tilde{\mathbf{s}})$ satisfy \eqref{eq:manager.Nash.KKT} with $\tilde{\lambda} = p(\tilde{\boldsymbol{\theta}}_d, \tilde{\boldsymbol{\theta}}_s)$.

{\bf Step 5. Uniqueness of the Nash Equilibrium.} 
We have shown that all Nash equilibria yield a unique market allocation. Uniqueness of the Nash equilibrium follows from the fact that the transformation from $\left( \boldsymbol{\theta}_d, \boldsymbol{\theta}_s\right)$ to $\left(\mathbf{d}, \mathbf{s}, \lambda\right)$ is one-to-one.

\section{Proof of Theorem \ref{prop:PoA}.}
{\bf Step 1. Bounding the Price Markups} To derive the upper bound on the Lerner index we note that at the Nash equilibrium there exists at least one firm such that
$S_i\left(p(\tilde{\boldsymbol{\theta}}_d, \tilde{\boldsymbol{\theta}}_s)\right) < \kappa_0$ or $\tilde{\theta}_s^i > 0$. Therefore, 
{\small {\begin{multline}
		\begin{split}
		p(\tilde{\boldsymbol{\theta}}_d, \tilde{\boldsymbol{\theta}}_s) & \leq \left(1+ \dfrac{S\left(\tilde{\theta}_s^i,p(\tilde{\boldsymbol{\theta}}_d, \tilde{\boldsymbol{\theta}}_s)\right)}{\zeta - \kappa_0}  \right) \dfrac{\partial C_i\left( S_i\left(\tilde{\theta}_s^i,p(\tilde{\boldsymbol{\theta}}_d, \tilde{\boldsymbol{\theta}}_s)\right) \right)}{\partial s_i} \leq \left( 1+ \dfrac{\kappa_0}{\zeta - \kappa_0}\right) \dfrac{\partial C_i \left( S_i\left(\tilde{\theta}_s^i,p(\tilde{\boldsymbol{\theta}}_d, \tilde{\boldsymbol{\theta}}_s)\right)\right)}{\partial s_i}\\ &\leq \dfrac{\zeta}{\zeta - \kappa_0} \max_{i}\left\lbrace\dfrac{\partial C_i \left( S_i\left(\tilde{\theta}_s^i,p(\tilde{\boldsymbol{\theta}}_d, \tilde{\boldsymbol{\theta}}_s)\right)\right)}{\partial s_i}\right\rbrace.\label{eq:Li.condition}
		\end{split}
		\end{multline}}}
Utilizing \eqref{eq:Li.condition} and substituting in the expression of $\LI(\tilde{\boldsymbol{\theta}}_d, \tilde{\boldsymbol{\theta}}_s)$ yields the bound in \eqref{eq:Lerner.bound}.

{\bf Step 2. Bounding the Social Welfare.} 
Let $\mathbf{x} = (\mathbf{d}, \mathbf{s})$ and $\tilde{\mathbf{x}} = (\tilde{\mathbf{d}}, \tilde{\mathbf{s}})$. In this step we aim to bound the social welfare at the Nash equilibrium, i.e., ${\mathbb{S}}(\tilde{\mathbf{x}})$. Specifically,

{\small{
		\begin{subequations}
			\begin{align}
			\mathbb{S}(\tilde{\mathbf{x}}) & \geq \mathbb{S}(\tilde{\mathbf{x}}) + \sum_{i=1}^{M+N}\dfrac{\partial\tilde{\mathbb{S}}_i(\tilde{x}_i)}{\partial x_i} (x_i^*-\tilde{x}_i) \label{eq:step1}\\
			&= \mathbb{S}(\tilde{\mathbf{x}}) + \left\lbrace \sum_{i=1}^{M} \dfrac{\partial \tilde{U}_i(\tilde{d}_i)}{\partial d_i}(d_i^* - \tilde{d}_i) - \sum_{i=1}^{N} \dfrac{\partial \tilde{C}_i(\tilde{s}_i)}{\partial s_i}(s_i^*-\tilde{s}_i) \right\rbrace\\
			&= \mathbb{S}(\tilde{\mathbf{x}}) + \sum_{i=1}^{M} \left(1 - \dfrac{\tilde{d}_i}{\zeta+d_0}\right) \dfrac{\partial{U}_i(\tilde{d}_i)}{\partial d_i}(d_i^* - \tilde{d}_i) \notag \\
			& - \sum_{i=1}^{N} \left(1 + \dfrac{\tilde{s}_i}{\zeta-\kappa_0} \right)\dfrac{\partial{C}_i(\tilde{s}_i)}{\partial s_i}(s_i^*-\tilde{s}_i) \label{eq:step3}\\
			& \geq \mathbb{S}(\tilde{\mathbf{x}}) + \sum_{i=1}^{M} \left(1 - \dfrac{\tilde{d}_i}{\zeta+d_0}\right) \left(U_i(d_i^*) - U_i(\tilde{d}_i)\right) \notag\\
			& - \sum_{i=1}^{N} \left(1 + \dfrac{\tilde{s}_i}{\zeta-\kappa_0} \right) \left(C_i(s_i^*) - C_i(\tilde{s}_i)\right) \label{eq:step4}\\
			& \geq \sum_{i=1}^{M}U_i(\tilde{d}_i) - \sum_{i=1}^{N}C_i(\tilde{s}_i) + \sum_{i=1}^{M} \left(1 - \dfrac{\tilde{d}_i}{d_i^*}\right) \left(U_i(d_i^*) - U_i(\tilde{d}_i)\right) \notag \\
			&- \left(1 + \dfrac{\kappa_0}{\zeta-\kappa_0}\right) \sum_{i=1}^{N}\left(C_i(s_i^*) - C_i(\tilde{s}_i)\right) \label{eq:step5}\\
			& \geq \sum_{i=1}^{M} \left( \left(\dfrac{\tilde{d}_i}{d_i^*}\right)^2 + 1 - \dfrac{\tilde{d}_i}{d_i^*} \right)U_i(d_i^*) - \left(\dfrac{\zeta}{\zeta-\kappa_0}\right)\sum_{i=1}^{N}C_i(s_i^*) \label{eq:step6}\\
			&\geq \dfrac{3}{4}\sum_{i=1}^{M} U_i(d_i^*) - \left(\dfrac{\zeta}{\zeta-\kappa_0}\right)\sum_{i=1}^{N}C_i(s_i^*).
			\end{align}
		\end{subequations}}}
		Inequality \eqref{eq:step1} follows from the optimality conditions of \eqref{eq:market.Nash.Problem} while \eqref{eq:step3} from the definitions of $\tilde{U}_i$ and $\tilde{C}_i$. Inequality \eqref{eq:step4} follows from concavity of $U_i(d_i)$ and convexity of $C_i(s_i)$. Step \eqref{eq:step5} follows from the fact that $d_i^* < \zeta + d_0$ for every $i \in \mathcal{M}$ and $\tilde{s}_i\leq \kappa_0$ for every $i \in \mathcal{N}$. Inequality \eqref{eq:step6} follows from concavity of $U_i(d_i)$ and that 
		{\small{
				\begin{equation*}
				U_i\left(\left(1-\dfrac{\tilde{d}_i - d_0}{d_i^* - d_0}\right) d_0 + \dfrac{\tilde{d}_i - d_0}{d_i^* - d_0}d_i^*\right)\geq \left(1-\dfrac{\tilde{d}_i - d_0}{d_i^* - d_0}\right) U_i(d_0) + \dfrac{\tilde{d}_i - d_0}{d_i^* - d_0} U_i(d_i^*) \Rightarrow U_i(\tilde{d}_i) \gtrsim \dfrac{\tilde{d}_i}{d_i^*}U_i(d_i^*).
				\end{equation*}}}
		The last inequality follows from minimizing the expression $y^2 - y + 1$, which is minimized for $y^* = 1/2$, where $y=\tilde{d}_i/d_i^*$. Finally, note that $\left(\dfrac{\zeta}{\zeta-\kappa_0}\right)$ is a decreasing function of $\zeta$. Hence, when $\zeta \in \left\lbrack4\kappa_0, \infty\right)$ the highest value of $\left(\dfrac{\zeta}{\zeta-\kappa_0}\right)$ is 4/3. 

%%%%%%%%%%%%%%%%%
\end{document}

%% file: header.tex
\newcommand{\beq}{\begin{equation}}
\newcommand{\eeq}{\end{equation}}
\newcommand{\beqa}{\begin{eqnarray}}
\newcommand{\eeqa}{\end{eqnarray}}
\newcommand{\beqan}{\begin{eqnarray*}}
\newcommand{\eeqan}{\end{eqnarray*}}

\newcommand{\Rset}{\mathds{R}}

\newcounter{l1}
\newcounter{l2}
\newcounter{l3}
\newcommand{\bdotlist}{\begin{list}{$\bullet$}{}}
\newcommand{\bboxlist}{\begin{list}{$\Box$}{}}
\newcommand{\bbboxlist}{\begin{list}{\raisebox{.005in}{{\tiny
$\blacksquare$ \ \ }}}{}}
\newcommand{\bdashlist}{\begin{list}{$-$}{} }
\newcommand{\blist}{\begin{list}{}{} }
\newcommand{\barablist}{\begin{list}{\arabic{l1}}{\usecounter{l1}}}
\newcommand{\balphlist}{\begin{list}{(\alph{l2})}{\usecounter{l2}}}
\newcommand{\bAlphlist}{\begin{list}{\Alph{l2}.}{\usecounter{l2}}}
\newcommand{\bdiamlist}{\begin{list}{$\diamond$}{}}
\newcommand{\bromalist}{\begin{list}{(\roman{l3})}{\usecounter{l3}}}

\newtheorem{theorem}{Theorem}
\newtheorem{lemma}{Lemma}